\documentclass[prl,superscriptaddress,twocolumn,showpacs,preprintnumbers,amsmath,amssymb]{revtex4}
\usepackage{hyperref}
\usepackage{subfigure}
\usepackage{epsfig}
\usepackage{dcolumn}
\usepackage{bm}
\usepackage{graphicx}
\usepackage{endnotes}
\usepackage[sort&compress]{natbib}
\include{include}

\usepackage{atlasphysics}

\newcommand\ztoll{$Z\to\ell\ell$}
\newcommand\htolvlv{$H\to W^+W^-\to\ell^+\nu\ell^-\bar{\nu}$}
\newcommand\htollll{$H\to ZZ\to\ell^+\ell^-\ell^+\ell^-$}
\newcommand\htollvv{$H\to ZZ\to\ell^+\ell^-\nu\bar{\nu}$}
\newcommand\zztollvv{$ZZ\to\ell^+\ell^-\nu\bar{\nu}$}
\newcommand\htollqq{$H\to ZZ\to\ell^+\ell^- q \bar{q}$}

\newcommand\totallumi{1.04~${\rm fb}^{-1}$}

\def\met{\ensuremath{E_{\mathrm{T}}^{\mathrm{miss}}}}

\begin{document}
\title{\vspace*{1cm} Search for a Standard Model Higgs boson in the \htollvv\ decay channel with the ATLAS detector}
\preprint{CERN-PH-EP-2011-142}
\preprint{Submitted to Physical Review Letters}
\author{The ATLAS Collaboration}
\affiliation{}
\begin{abstract} 
A search for a heavy Standard Model Higgs~boson decaying via
\htollvv, where $\ell=e,\mu$, is presented.  
It is based on proton-proton collision data at $\sqrt{s}=7\tev$, collected by the ATLAS experiment at the LHC
in the first half of 2011 and corresponding to an integrated luminosity of \totallumi. The data are compared 
to the expected Standard Model backgrounds. The data and the background expectations are found to be in 
agreement and upper limits are placed on the Higgs~boson production cross section over the entire mass window considered; 
in particular the production of a Standard Model Higgs boson is excluded in the region  $340 < m_{\mathrm{H}} < 450$~GeV at the 95\% confidence level. 
\end{abstract}
\maketitle
\normalsize

The search for the Standard Model (SM)
Higgs boson~\cite{prl_13_321,Higgs:1964pj,Guralnik:1964eu}
is one of the most important aspects of the Large Hadron Collider (LHC) physics program. 
Direct searches at the CERN LEP $e^+e^-$ collider 
have set a lower limit of
$114.4\gev$ on the Higgs~boson mass, $m_{\mathrm{H}}$, 
at 95\% confidence level~\cite{lepew}.
Searches by the CDF and D0 experiments at the
Fermilab Tevatron $p\bar{p}$ collider
have explored the mass range up to $200\gev$ and
exclude the additional region
$156<m_{\mathrm{H}}<177\gev$~\cite{tevhiggsspring2011}. 
For $m_{\mathrm{H}}$ greater than twice the $Z$~boson mass, $m_{\mathrm{Z}}$, a significant 
fraction of Higgs~bosons 
decay to two $Z$~bosons.
The \zztollvv\ decay channel offers a substantial branching fraction in combination with a good separation from potential 
background processes owing to the high transverse momentum, $p_{\mathrm{T}}$, of the electron or muon pair from the leptonic $Z$ decay and 
the high missing transverse momentum, \met, from the $Z$ decaying to neutrinos. 

The first cross section limits for a SM Higgs~boson in the mass region $200 < m_{\mathrm{H}} < 600$~GeV were set by the ATLAS and CMS
collaborations in Refs.~\cite{atlas2010,cms2010}. This letter extends the \htollvv\ results therein, with a 30-fold increase in the 
integrated luminosity, as well as a significant improvement in the event reconstruction  
and background rejection. 

The data sample considered in this search was recorded by the ATLAS experiment during the first half 
of the 2011 LHC run at a center-of-mass energy $\sqrt{s}=7$~TeV. 
The integrated luminosity of the data sample, considering only data-taking periods where all relevant 
detector subsystems were 
operational, is \totallumi.

The ATLAS detector has been described elsewhere~\cite{atlas}. Simulated signal and background event samples are produced with Monte Carlo (MC) event generators, passed 
through a full GEANT4~\cite{GEANT4} simulation of the ATLAS detector~\cite{:2010wqa} and reconstructed with the same reconstruction 
software as the data. 

\htollvv\ ($\ell=e,\mu,\tau$) events are modelled using the POWHEG~\cite{powheg4,powheg5} event generator, which includes matrix elements for the gluon fusion and the vector-boson fusion production mechanisms of the Higgs~boson up to 
next-to-leading order. POWHEG is interfaced to PYTHIA~\cite{pythia} for the modelling of parton showers. The Higgs boson \pt~spectrum is reweighted to the calculation of Ref.~\cite{Bozzi:2005wk}, which provides QCD corrections up to next-to-leading order and QCD soft-gluon resummations up to next-to-next-to-leading logarithms. An alternative sample of signal events is produced using the PYTHIA event generator, 
which includes only leading order matrix elements. In both cases PHOTOS~\cite{Golonka:2005pn} is used to model
final-state radiation and TAUOLA~\cite{Was:2000st} for the simulation of $\tau$ decays. 

\htollll\ and \htollqq\ samples are also simulated using the same generators as for the \htollvv\ samples, whilst \htolvlv\ events 
are produced using the MC@NLO generator~\cite{mcatnlo}, interfaced to HERWIG~\cite{herwig} and JIMMY~\cite{jimmy} 
in the gluon fusion channel and the SHERPA~\cite{Gleisberg:2008ta} generator in the vector-boson fusion channel. These channels  
contribute to the signal yield and are considered as part of the signal. 
In particular, \htolvlv\ decays contribute as much as 77\% to the signal expectation after the full selection for $m_{\mathrm{H}}=200$~GeV decreasing to 13\% at $m_{\mathrm{H}}=300$~GeV.
Independence of the analysis with respect to other ATLAS Higgs boson searches~\cite{htoww2011,hto4l2011,Aad:2011ec} is ensured through mutually exclusive selection requirements on the dilepton invariant mass, the number of leptons or the event missing transverse momentum.

The cross sections for Higgs~boson production, the associated branching fractions~\cite{Djouadi:1997yw}, as well as their uncertainties, 
are compiled in Ref.~\cite{LHCHiggsCrossSectionWorkingGroup:2011ti}. They correspond to next-to-next-to-leading order 
in QCD for the gluon fusion~\cite{Harlander:2002wh,Anastasiou:2002yz,Ravindran:2003um,Anastasiou:2008tj,deFlorian:2009hc,Baglio:2010ae} 
and the vector boson fusion~\cite{Bolzoni:2010xr} processes.
In addition, QCD soft-gluon resummations up to next-to-next-to-leading logarithms 
are  available for the gluon fusion process~\cite{Catani:2003zt}, while next-to-leading order electroweak corrections are applied to both the 
gluon fusion~\cite{Aglietti:2004nj,Actis:2008ug} and the vector boson fusion~\cite{Ciccolini:2007jr,Ciccolini:2007ec} 
processes. 
These cross section calculations do not account for the width of the 
Higgs~boson, which is implemented through an ad-hoc Breit-Wigner line shape 
applied at the event generator level. Recent studies~\cite{LHCHiggsCrossSectionWorkingGroup:2011ti,Anastasiou:2011pi} 
have indicated 
that effects due to off-shell Higgs~boson production and interference 
with other SM processes may become sizeable at the highest masses 
($m_{\mathrm{H}}>400$~GeV) considered in this search. In the absence of a full 
calculation, a conservative estimate of the possible size of such effects was made and the impact 
on the obtained limits in this channel was found to be less than 
2\% for $m_{\mathrm{H}}=400$~GeV growing to about 25\% at $m_{\mathrm{H}}=600$~GeV. 

Different event generators are chosen to model a range of important background processes. The ALPGEN
generator~\cite{alpgen} interfaced with HERWIG for parton showers and
hadronisation is used to simulate $W/Z+$jets backgrounds. MC@NLO, interfaced to
HERWIG and JIMMY, is used for the production of 
top-pair, single top and diboson ($WW$, $WZ$ and $ZZ$) backgrounds. PYTHIA is used to simulate 
$b\bar{b}$ and $c\bar{c}$ samples as well as alternative samples for the $Z$ and $ZZ$ backgrounds. 
All simulated background samples are scaled to the highest available precision calculations for the relevant process. 
An overview of the used predictions and their uncertainties is given in Ref.~\cite{zjetsConf}.

Data used for the search in the electron and muon channels were collected primarily using single lepton triggers
with $p_{\mathrm{T}}$ thresholds of 20 and 18~GeV respectively. 
The expected trigger efficiency is close to 100\% in the electron channel and about 95\% in the muon channel for 
signal events passing all the selection criteria described below

Electron candidates are reconstructed from electromagnetic calorimeter clusters, with shapes consistent with those 
expected from electromagnetic showers, matched to tracks reconstructed in the inner detector. 
Details of the electron reconstruction and identification can be found in Ref.~\cite{Aad:2010yt}.
The electron candidates are required to pass the standard ATLAS ``medium'' selection criteria and have $p_{\mathrm{T}}>20$~GeV and 
pseudorapidity $|\eta|<2.47$.

Muons are identified by reconstructing tracks in the muon spectrometer. These tracks are then 
extrapolated back to the beam line to find a matching inner detector track. 
Details of muon reconstruction and identification can be found in Ref.~\cite{Aad:2010yt}.
Only muons with $p_{\mathrm{T}}>20$~GeV and $|\eta| < 2.5$ are considered.

Jets are used in this analysis to reject backgrounds from events with heavy quark decays or 
from events with fake \met\ due to mis-measured jets. For this purpose 
jets are reconstructed from 
clusters of energy deposits in the calorimeters using the 
anti-$k_{\mathrm{t}}$ algorithm~\cite{AntiKt} with a radius parameter $R=0.4$. Only jets with $p_{\mathrm{T}}>25$~GeV and  $|\eta|<2.5$ are considered.

To remove leptons associated with jets, such as those originating from semi-leptonic decays of 
$b$ hadrons, leptons are not considered in the analysis if the sum of inner detector track momenta in a cone 
$\Delta R < 0.2$ around the lepton direction is greater than $10\%$ of the $p_{\mathrm{T}}$ of the lepton itself or 
if the lepton is within a distance $\Delta R<0.4$ of the nearest jet.

The missing transverse momentum is measured as the (negative) vectorial sum of the transverse momenta of all 
clusters in the calorimeters within $|\eta|<4.5$ 
and all selected muons in the event. Calorimeter deposits associated with muons are subtracted to avoid double counting.

\begin{figure}[htb]
\includegraphics[width=0.89\columnwidth]{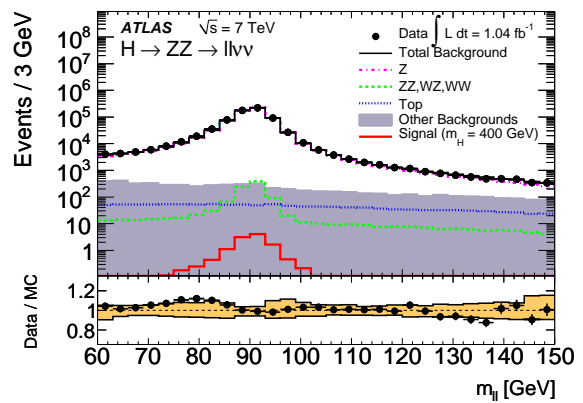}
  \caption{The dilepton invariant mass distribution for events with exactly two oppositely charged electrons or muons. The inset at the bottom of the figure shows the ratio between the data and the combined background expectations as well as a band corresponding to the combined systematic uncertainties of the analysis.\label{fig:htollvvcontrol}}
\end{figure}

Events are required to contain a reconstructed 
primary vertex formed from at least 3 tracks and exactly two oppositely charged electrons or muons, consistent with originating from the primary vertex. 
The dilepton mass distribution is shown in Fig.~\ref{fig:htollvvcontrol}. 
Inclusive $Z$~boson production is the dominant background at this stage of the analysis. 
To suppress backgrounds from top, $W$, and QCD multijet production, the 
dilepton invariant mass, $m_{\ell\ell}$, is required to satisfy $|m_{\mathrm{Z}}-m_{\ell\ell}| < 15$~GeV. 

\begin{figure*}[thb]
\includegraphics[width=0.43\textwidth]{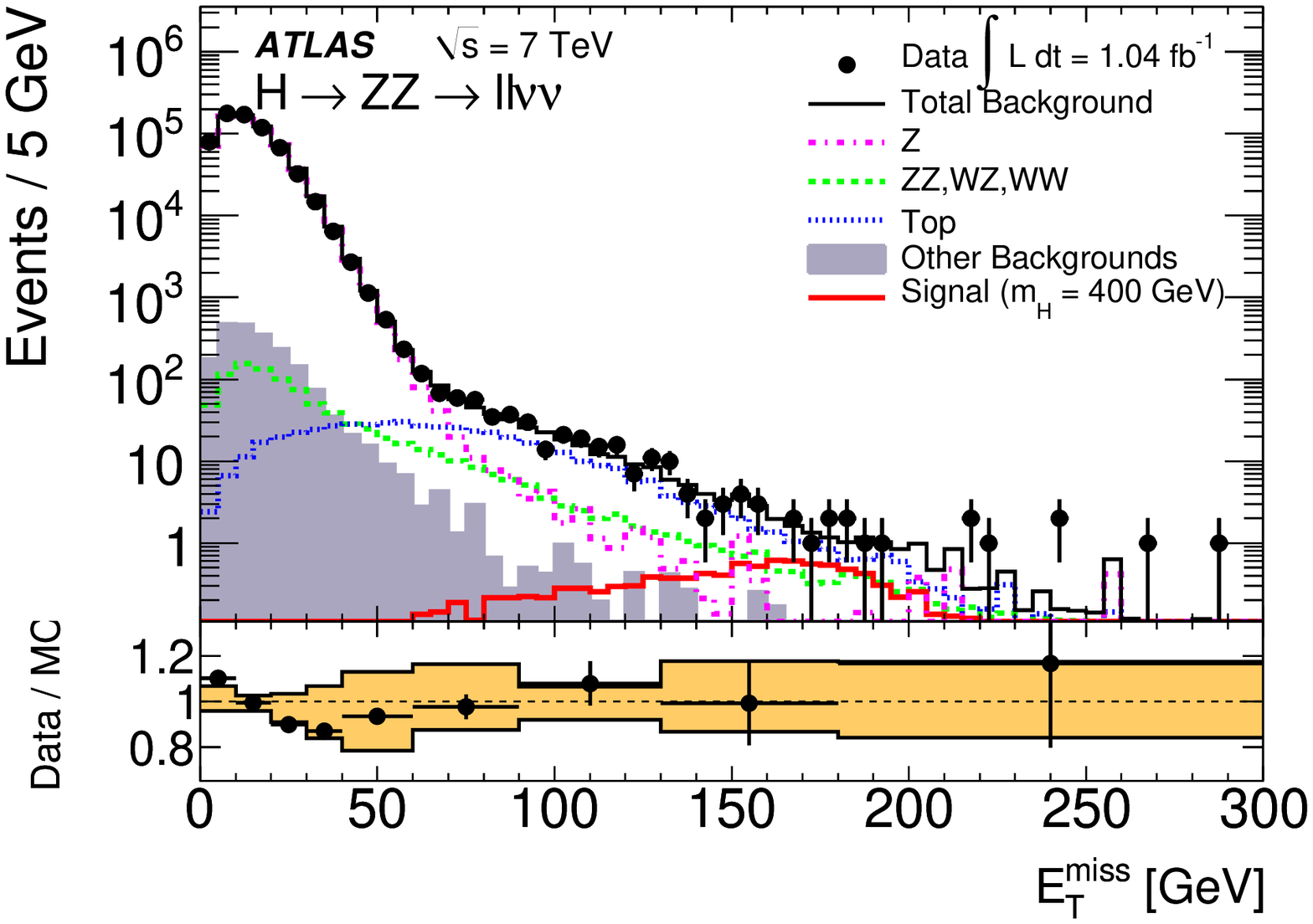}
\includegraphics[width=0.43\textwidth]{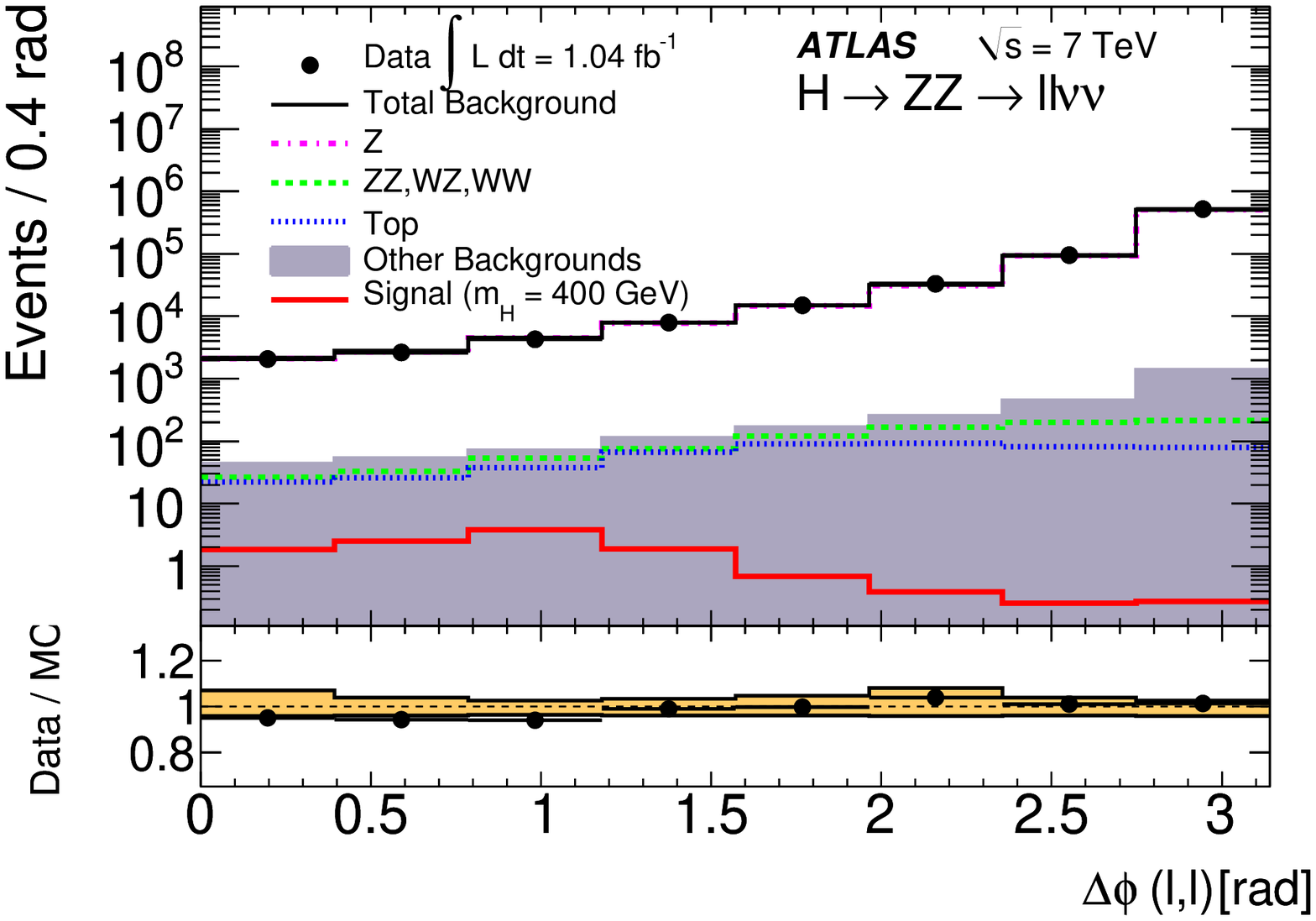}
\caption{The \met\ (left) and $\Delta\phi(\ell,\ell)$ (right) distributions for events with exactly two oppositely charged electrons or muons inside the $Z$ mass window. The insets at the bottom show the ratio between data and the combined background expectations as well as a band corresponding to the combined systematic uncertainties of the analysis.\label{fig:htollvvcontrol2}}
\end{figure*}

To reduce the background from events with fake \met\ due to mis-measured jets, events are rejected if the azimuthal angle between the missing transverse 
momentum vector, $\vec{p}_\mathrm{T}^{\rm\ miss}$, 
and the leading jet in the event satisfies $\Delta\phi(\vec{p}_\mathrm{T}^{\rm\ miss},\vec{p}_\mathrm{T}^{\rm jet}) < 0.3$. To reduce the background from top~quark production, 
events with one or more $b$-tagged jets are rejected, where the $b$-tagging is based on a single discriminant combining information from both the impact parameter with respect to the primary vertex of tracks associated to the jet and the presence of displaced secondary vertices associated to the jet's tracks. 
The chosen cut achieves an efficiency of about 70\% for identifying real $b$-jets, with a light-quark jet rejection of about~80~\cite{btaghigh}. 

To exploit the mass dependent kinematic features of \htollvv\ production, the search is subdivided into a low Higgs~boson mass ($m_{\mathrm{H}}<280$~GeV) and a high Higgs~boson mass ($m_{\mathrm{H}}\geq280$~GeV) search region, where dedicated cuts are applied to two important discriminating variables used to reduce the background contributions: \met\ and the azimuthal angle between the two leptons, $\Delta\phi(\ell,\ell)$. 
Figure~\ref{fig:htollvvcontrol2} shows the distributions of these variables after the application of the $m_{\ell\ell}$ window cut.
Since inclusive $Z$ production gives rise to a steeply falling \met\ distribution, systematic uncertainties on the \met\ reconstruction are particularly important to estimate this background correctly. The dominant contributions to the \met\ uncertainty come from the knowledge of the jet energy scale and the modelling of inclusive $Z$ production. Figure~\ref{fig:htollvvcontrol2} shows that a good agreement within systematic uncertainties is observed between data and the combined background expectation.  
In the low $m_{\mathrm{H}}$  region, events are required to satisfy $\met>66$~GeV, whilst in the high $m_{\mathrm{H}}$ region the requirement is  $\met>82$~GeV. These cuts reduce significantly the backgrounds from processes with no or modest genuine missing transverse momentum originating from unobserved neutrinos.

The boost of the $Z$~bosons originating from a Higgs~boson decay increases with $m_{\mathrm{H}}$, thus reducing the expected opening angle between the leptons. 
In the low $m_{\mathrm{H}}$ region this boost is expected to be modest and a cut $1 < \Delta\phi(\ell,\ell) < 2.64$ is applied. In the high $m_{\mathrm{H}}$ region an upper limit $\Delta\phi(\ell,\ell) < 2.25$ is required. 

Finally, in the high $m_{\mathrm{H}}$ region, events are also rejected if the azimuthal angle between the missing transverse momentum vector  
and the direction of the \ztoll~boson candidate is $\Delta\phi(\vec{p}_\mathrm{T}^{\rm\ miss},\vec{p}_\mathrm{T}^{\ell\ell}) < 1$.
The efficiency of the event selection is very similar in the electron and muon channels, ranging from 3\% for $m_{\mathrm{H}}=200$~GeV to about 48\% for $m_{\mathrm{H}}=600$~GeV.

SM pair production of $Z$~bosons has a final state identical to the signal, and is therefore expected to survive most of the
applied selection criteria and form a continuum in the transverse mass distribution (defined below). 
The normalisation for this background is obtained from a calculation including 
next-to-leading order terms~\cite{Campbell:1999ah} with an additional 6\% term to account for missing quark-box diagrams 
($gg\to  ZZ$)~\cite{Binoth:2008pr}. A 11\% normalisation uncertainty is assigned to this background, estimated from scale, PDF and model uncertainties. $WW$ and $WZ$ backgrounds are normalised in a similar way.

The background from inclusive $Z$ production is derived from MC, after checking that the simulation describes well the 
data in samples selected by requiring the presence of a lepton pair.
The  background from top events is also taken from the
MC prediction. This prediction is verified to agree
with data, within systematic uncertainties, in two independent
control samples: the first one requires at least one identified $b$-jet,
while the second selects events containing electron-muon pairs.

Additional backgrounds can arise from QCD multijet events or inclusive $W$ production due to heavy flavour 
decays or jets faking leptons. 
The normalisation of the $W$ background is obtained from the ratio between data and MC in  control 
samples of like-sign electron-electron and electron-muon events with high \met. 
The QCD multijet background in the electron channel is determined using a data sample based on a 
loosened electron selection, thus dominated by jets; this sample is scaled to describe the tails of the $m_{\ell\ell}$ distribution. 
In the muon channel, the background from heavy flavour decays is studied using simulation, whereas other muon sources
from multijet events are constrained using a sample of like-sign muon pairs in data. In both cases the background is found to 
be negligible.

The signal efficiencies and overall background expectations are similar in the electron and the muon channels, therefore only combined results are presented.  
The numbers of candidate \htollvv\ events selected in data and the expected yields from signal and background processes are 
shown in Table~\ref{tab:htollvv_yield}. 
\begin{table}[htb]
\caption{
The expected number of background and signal events for the Higgs~boson search in the \htollvv\ channel, along with the observed numbers of candidates in data, for an integrated luminosity of \totallumi.
The quoted uncertainties are statistical and systematic respectively. 
Signal to background ratios are also given for various masses (see text).
\label{tab:htollvv_yield}}
{\small
\begin{tabular}{crrcc}\hline\hline
 Source      &  low $m_{\mathrm{H}}$ search          &   high $m_{\mathrm{H}}$ search    &  & \\  \hline
  $Z$        &  $19.1\pm 2.6 \pm 0.9$     &   $6.0\pm 1.4 \pm 1.8$ \\
  $W$        &  $8.5\pm 2.3 \pm 8.5$      &   $3.1\pm 1.0 \pm 3.1$ \\
  top        &  $29.9\pm 1.3 \pm 6.0$     &   $14.9\pm 0.8 \pm 3.1$ \\
  multijet   &  $0.4\pm 0.4 \pm 0.2$      &   $0.0\pm 0.0 \pm 0.0$ \\
  $ZZ$       &  $17.6\pm 0.4 \pm 2.1$     &   $14.7\pm 0.4 \pm 1.7$ \\
  $WZ$       &  $16.7\pm 0.6 \pm 2.0$     &   $12.1\pm 0.5 \pm 1.4$\\
  $WW$       &  $12.4\pm 0.4 \pm 1.5$     &   $4.6\pm 0.3 \pm 0.5$   \\ \hline
  Total      &  $104.6 \pm 3.8  \pm 16.0$ &   $55.3 \pm 2.0  \pm 7.8$ \\
  Data       &         85                 &          47     \\ \hline 
  $m_{\mathrm{H}}$ [GeV]      & \multicolumn{2}{c}{Signal expectation }   & $s/b$ \\ 
 \hline
  200  &    $5.0\pm 0.1 \pm 0.9$   &       &7\%  \\
  300  &        &    $10.2\pm 0.2 \pm 1.8$ &22\% \\
  400  &        &     $10.0\pm 0.2 \pm 1.7$&52\% \\
  500  &        &     $4.5\pm 0.1 \pm 0.8$ &57\% \\
  600  &        &    $1.8\pm 0.0 \pm 0.3$  &60\% \\ \hline\hline
\end{tabular}}
\end{table}

The systematic uncertainties include experimental uncertainties related to 
the selection and calibration of electrons, muons, jets and $b$-jets, which are also explicitly propagated to the \met\ calculation. 
Shape uncertainties for the signal and for the single $Z$ and $ZZ$ backgrounds are estimated using PYTHIA as an alternative 
MC generator.

Normalisation uncertainties for signal (gluon fusion {\scriptsize$^{+14\%}_{-10\%}$} and VBF 4\%) and 
diboson backgrounds (11\%) are obtained from theory~\cite{LHCHiggsCrossSectionWorkingGroup:2011ti}; uncertainties for the inclusive $Z$~boson production (2.5\%), top~quark production (9\%), 
inclusive $W$~boson production (100\%) and 
QCD multijet production in the electron channel (50\%) are estimated from data.
A 3.7\% luminosity uncertainty~\cite{lumiupdatesummer} is included for those processes for which the normalisation is not obtained from the data. 
The dominant systematic uncertainties in the analysis are the \met\ uncertainties 
for the $Z$ background, the $b$-tagging uncertainty for the top background and 
the normalisation uncertainties for the signal and the $W$ and diboson backgrounds.

After the event selection, the Higgs~boson search is performed by 
looking for an excess of data over the 
SM background expectation in the transverse mass distribution of the selected $ee\nu\nu$ and 
$\mu\mu\nu\nu$ events. The transverse mass is calculated from the lepton pair and the $\vec{p}_\mathrm{T}^{\rm\ miss}$ vector as:

{\scriptsize \[ m_{\mathrm{T}}^2 \equiv \left[\sqrt{m_{\mathrm{Z}}^2+|\vec{p}_\mathrm{T}^{\ \ell\ell}|^2}+\sqrt{m_{\mathrm{Z}}^2+|\vec{p}_\mathrm{T}^{\rm\ miss}|^2}\right]^2-\left[\vec{p}_\mathrm{T}^{\ \ell\ell}+\vec{p}_\mathrm{T}^{\rm\ miss}\right]^2.\]}

\begin{figure}[htb]
\includegraphics[width=0.89\columnwidth]{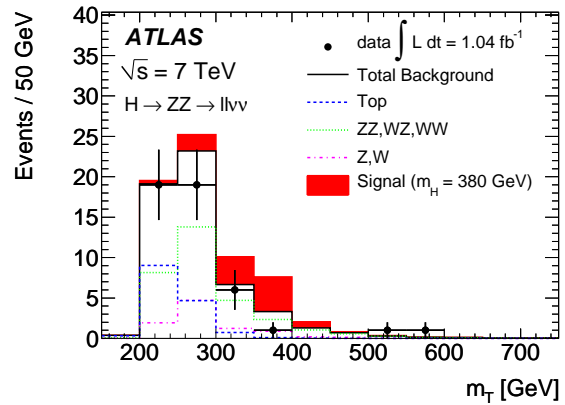}
\caption{\label{fig:htollvv_mass}The transverse mass distribution of \htollvv\ candidates in the high $m_{\mathrm{H}}$ 
search region for the data (dots), the expected backgrounds (histograms) and a Higgs boson of mass 380~GeV (filled histogram). The electron and muon channels are combined. }           
\end{figure}

Figure~\ref{fig:htollvv_mass} shows the $m_{\mathrm{T}}$ distribution in the high $m_{\mathrm{H}}$ search region.  
Signal to background ratios for different $m_{\mathrm{H}}$ values, determined in a $m_{\mathrm{T}}$ window defined to enclose 95\% of the corresponding signal events, are listed in Table~\ref{tab:htollvv_yield}.

The number and  distribution of candidate \htollvv\ events observed in the data agree with the expected backgrounds within the uncertainties, with no 
indication of an excess.
Upper limits are set on the Higgs~boson production cross section relative to its predicted SM value as a 
function of $m_{\mathrm{H}}$. The limits are extracted from a maximum likelihood fit to the $m_T$ distribution following the $CL_s$ modified frequentist formalism with the profile likelihood test statistic~\cite{Read:2002hq,asymptotics}. All systematic uncertainties are taken into account.

\begin{figure}[htb]
  \includegraphics[width=0.98\columnwidth]{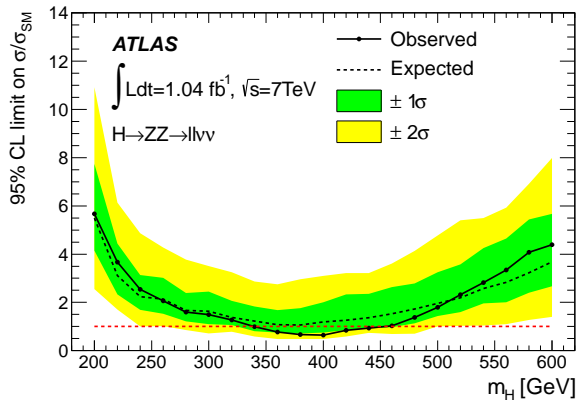}
  \caption{Observed and expected 95\% confidence level upper limits on the Higgs boson production cross~section divided by the SM prediction. The green and yellow bands indicate the $\pm 1\sigma$\, and $\pm 2 \sigma$\, fluctuations, respectively, around the median sensitivity. The limits are based on \totallumi\ of data at $\sqrt{s}=$7~TeV. \label{fig:limits_llnunu_sig}}
\end{figure}

Figure~\ref{fig:limits_llnunu_sig} shows the expected and observed limits at the 95\% confidence level. The expected
limit is lowest around  $m_{\mathrm{H}}=380$~GeV where it is 1.1 times the SM Higgs boson cross~section. Fluctuations in the
background can lead to better or worse expected limits. Over the entire mass range the observed limits agree with the 
expectations within the  $\pm 2 \sigma$\, band. A SM Higgs boson in the
range 340~GeV~$<m_{\mathrm{H}}<$~450~GeV is excluded at the 95\% confidence level

In summary, results of a search for a heavy SM Higgs~boson with a mass in the range $200 < m_{\mathrm{H}} < 600$~GeV decaying to
 $ZZ\to\ell^+\ell^-\nu\bar{\nu}$ have been presented. These results are based on a data sample corresponding
to an integrated luminosity of \totallumi, recorded with the ATLAS detector at the LHC. No evidence for a signal is observed and 
cross section limits are placed over the entire mass range, excluding the production of a SM Higgs~boson 
in the region $340 < m_{\mathrm{H}} < 450$~GeV at the 95\% confidence level. 

\quad
\begin{acknowledgments}
We thank CERN for the very successful operation of the LHC, as well as the
support staff from our institutions without whom ATLAS could not be
operated efficiently.

We acknowledge the support of ANPCyT, Argentina; YerPhI, Armenia; ARC,
Australia; BMWF, Austria; ANAS, Azerbaijan; SSTC, Belarus; CNPq and FAPESP,
Brazil; NSERC, NRC and CFI, Canada; CERN; CONICYT, Chile; CAS, MOST and
NSFC, China; COLCIENCIAS, Colombia; MSMT CR, MPO CR and VSC CR, Czech
Republic; DNRF, DNSRC and Lundbeck Foundation, Denmark; ARTEMIS, European
Union; IN2P3-CNRS, CEA-DSM/IRFU, France; GNAS, Georgia; BMBF, DFG, HGF, MPG
and AvH Foundation, Germany; GSRT, Greece; ISF, MINERVA, GIF, DIP and
Benoziyo Center, Israel; INFN, Italy; MEXT and JSPS, Japan; CNRST, Morocco;
FOM and NWO, Netherlands; RCN, Norway; MNiSW, Poland; GRICES and FCT,
Portugal; MERYS (MECTS), Romania; MES of Russia and ROSATOM, Russian
Federation; JINR; MSTD, Serbia; MSSR, Slovakia; ARRS and MVZT, Slovenia;
DST/NRF, South Africa; MICINN, Spain; SRC and Wallenberg Foundation,
Sweden; SER, SNSF and Cantons of Bern and Geneva, Switzerland; NSC, Taiwan;
TAEK, Turkey; STFC, the Royal Society and Leverhulme Trust, United Kingdom;
DOE and NSF, United States of America.

The crucial computing support from all WLCG partners is acknowledged
gratefully, in particular from CERN and the ATLAS Tier-1 facilities at
TRIUMF (Canada), NDGF (Denmark, Norway, Sweden), CC-IN2P3 (France),
KIT/GridKA (Germany), INFN-CNAF (Italy), NL-T1 (Netherlands), PIC (Spain),
ASGC (Taiwan), RAL (UK) and BNL (USA) and in the Tier-2 facilities
worldwide.
\end{acknowledgments}

\bibliography{paper,Higgs_paper}
\onecolumngrid
\clearpage
\onecolumngrid
% ATLAS Collaboration author list for 09-JUL-2011
% Data extracted on 15-Aug-2011 for paperid 117
%\documentclass[11pt]{article}
%\usepackage{a4wide}\begin{document}
\begin{flushleft}
{\Large The ATLAS Collaboration}

\bigskip

G.~Aad$^{\rm 48}$,
B.~Abbott$^{\rm 111}$,
J.~Abdallah$^{\rm 11}$,
A.A.~Abdelalim$^{\rm 49}$,
A.~Abdesselam$^{\rm 118}$,
O.~Abdinov$^{\rm 10}$,
B.~Abi$^{\rm 112}$,
M.~Abolins$^{\rm 88}$,
H.~Abramowicz$^{\rm 153}$,
H.~Abreu$^{\rm 115}$
% ...
\bigskip

% ...
$^{10}$ Institute of Physics, Azerbaijan Academy of Sciences, Baku, Azerbaijan\\
$^{11}$ Institut de F\'isica d'Altes Energies and Departament de F\'isica de la Universitat Aut\`onoma  de Barcelona and ICREA, Barcelona, Spain\\
% ...
$^{48}$ Fakult\"{a}t f\"{u}r Mathematik und Physik, Albert-Ludwigs-Universit\"{a}t, Freiburg i.Br., Germany\\
$^{49}$ Section de Physique, Universit\'e de Gen\`eve, Geneva, Switzerland\\
% ...
$^{88}$ Department of Physics and Astronomy, Michigan State University, East Lansing MI, United States of America\\
% ...
$^{111}$ Homer L. Dodge Department of Physics and Astronomy, University of Oklahoma, Norman OK, United States of America\\
$^{112}$ Department of Physics, Oklahoma State University, Stillwater OK, United States of America\\
% ...
$^{115}$ LAL, Univ. Paris-Sud and CNRS/IN2P3, Orsay, France\\
% ...
$^{118}$ Department of Physics, Oxford University, Oxford, United Kingdom\\
% ...
$^{153}$ Raymond and Beverly Sackler School of Physics and Astronomy, Tel Aviv University, Tel Aviv, Israel\\
% ...
\end{flushleft}

%\end{document}

\end{document}